# TRAFFIC SHAPING AND HYSTERESIS MITIGATION USING DEEP REINFORCEMENT LEARNING IN A CONNECTED DRIVING ENVIRONMENT

**Rami Ammourah**
Civil and Environmental Engineering Department
University of Illinois at Urbana Champaign
Urbana, IL 61801
ramiaa2@illinois.edu

**Alireza Talebpour**
Civil and Environmental Engineering Department
University of Illinois At Urbana Champaign
Urbana, IL 61801
ataleb@illinois.edu

February 6, 2023

## ABSTRACT

This paper proposes a multi-agent deep reinforcement learning-based framework for traffic shaping, i.e., maintaining a desirable traffic flow given a certain traffic density. The proposed framework offers a key advantage over existing congestion management strategies which is the ability to mitigate hysteresis phenomena. In other words, unlike the majority of congestion management strategies that focus on breakdown prevention/delay, the proposed framework is still extremely effective after the breakdown formation. The proposed framework assumes partial connectivity embodied by the connected automated vehicles (CAVs) which share information amongst each other. Moreover, the framework only requires a basic level of autonomy (Level 1), which is defined by a one-dimensional longitudinal control performed by the automated vehicles. This framework is primarily built using a centralized training, centralized execution multi-agent deep reinforcement learning approach, where longitudinal control is defined by signals of acceleration or deceleration commands which are then executed by all agents uniformly (i.e. centralized execution). The model undertaken for training and testing of the framework is based on the well-known Double Deep Q-Learning algorithm (DDQN) which takes the average state of flow within the traffic stream as the model input (state) and outputs actions in the form of acceleration or deceleration values. We demonstrate the ability of the model to shape the state of traffic, mitigate the negative effects of hysteresis, and even improve traffic flow beyond its original level. This paper also identifies the minimum percentage of CAVs required to successfully shape the traffic under an assumption of uniformly distributed CAVs within the loop system. The framework illustrated in this work doesn't just show the theoretical applicability of reinforcement learning to tackle such challenges, but also proposes a realistic solution that only requires partial connectivity and continuous monitoring of the average speed of the system, which can be achieved using readily available sensors that measure the speeds of vehicles in reasonable proximity to the CAVs.

***Keywords*** Deep Reinforcement Learning, Automated and Connected Vehicles, Traffic Shaping, Congestion Management, Hysteresis Phenomena

# 1 Introduction

Roadway congestion has been a major challenge in urban environments for the past several decades, negatively impacting both commuters and the environment. Accordingly, traffic management and congestion mitigation strategies have been investigated extensively in the past few decades and various solutions have been proposed spanning both demand and supply sides (and often both), including congestion pricing [1, 2, 3], speed harmonization and variable message signs [4, 5, 6, 7, 8], ramp metering [9, 10, 11, 12, 13, 14], and High Occupancy Vehicle (HOV) lanes [15, 16].

Traffic Shaping and Hysteresis Mitigation Using Deep Reinforcement Learning in a Connected Driving EnvironmentUnfortunately, due to the complexity of the road environment and interactions among various roadway users, effective traffic management remains an open challenge for researchers and practitioners around the world.

While classical approaches to traffic management showed some success, they were mostly reactive methods and relied mostly on historical and/or low-resolution data (from infrastructure sensors). Recent developments in connected and automated vehicle (CAV) technology as well as in artificial intelligence have provided the opportunity to address many of these challenges and has opened the door to new research endeavors in congestion and traffic management. A good example of such efforts is a study by Elfar [17]. He used logistic regression, random forests, and neural networks to develop offline and online predictive models to predict short-term traffic congestion using vehicle trajectories available through connected vehicles. Moreover, researchers have started looking at the ability of reinforcement learning to create higher-performing models and utilized these predictions to implement a predictive speed harmonization system. Mohanty et al. [18] used Long Short-Term Memory (LSTM) networks to forecast neighborhood congestion from region-wide observations. Later the LSTM model was used to approximate optimal tolling required for congestion mitigation. On the other hand, Genser et al. [19] investigated dynamic optimal congestion pricing for the purpose of traffic management by the application of multi-layer neural networks. They utilized neural network models to predict generalized costs and derive optimal pricing functions. Other methods proposed in the literature include the use of Bayesian regression [20], Convolutional Neural Networks [21], and clustering [22].

Recently, the combination of reinforcement learning with CAVs has gained a lot of attention. Han et al. put forward a physics-informed reinforcement learning(RL)-based ramp metering strategy. They train the RL model using a combination of historic data and synthetic data generated from a traffic flow model to avoid falling into an inaccurate training environment. The strategy is applied to both local and coordinated ramp metering, and results show significant improvements to traffic performance and that it outperforms classical feedback-based ramp metering strategies [23]. Earlier implementations utilized reinforcement learning in its classical form by creating a Q-learning model and training it through actions and rewards [24, 25]. More recently, the emergence of deep reinforcement learning has also opened the door for further explorations with several studies adopting the Deep Q Network (DQN) and variations of it to tackle multiple problems in traffic management [26, 27, 28]. Walraven et al. [29] modeled the traffic flow optimization problem as a reinforcement learning problem. They obtain speed limit policies using Q-learning and use neural networks to improve the performance of their policy learning algorithm. Their method considers traffic predictions and outputs control signals proactively. Wen et al. [30] utilized a deep Q network (DQN) to address re-balancing needs that are critical for effective fleet management in shared mobility on-demand systems. They show that DQN performs effectively by reducing wait times for travelers and limiting the distance traveled by vehicles. In another study, Zhou et al. [31] used the Deep Deterministic Policy Gradient algorithm to create a reinforcement-learning-based car-following model for CAVs to obtain appropriate driving behavior to improve travel efficiency, fuel consumption, and safety at signalized intersections in real-time. Other works have utilized multi-agent reinforcement learning for the task of large-scale traffic signal control. Variations of cooperative deep reinforcement learning are used and they demonstrate the ability of their frameworks to alleviate congestion significantly [32, 33, 34].

Taking a closer look at recent works that adopt reinforcement learning as a technique to alleviate traffic congestion, none focus on improving traffic flow within the hysteresis region. For instance, Kreidieh et al. [35] focus on dissipating stop-and-go waves as they occur in either closed or open networks using a percentage of connected and automated vehicles that follow a reinforcement learning control strategy to achieve the task. (I.e. the strategy is successful in preventing stop-and-go traffic from accumulating and causing increased slowdowns, however, it remains a proactive strategy in the sense that it has no utility should congestion accumulate. Ha et al. similarly utilize graph convolutional networks and reinforcement learning to improve throughput at a bottlenecked segment [36]. On the other hand, other works such as that of Zhu et al. [37] look at car following behavior and leverage reinforcement learning to optimize driver performance in a natural driving environment (simulated by the NGSIM dataset) and do not focus on congestion mitigation.

A critical shortcoming of most existing literature is that they focus on preventing and/or delaying the breakdown formation. However, once congestion happens and breakdown occurs, almost no strategy is available to mitigate or eliminate congestion. In fact, it is a widely accepted notion that once breakdown happens, not much can be done to eliminate congestion and without any intervention, the recovery happens only through demand reduction. Unfortunately, in the fundamental diagram of traffic flow, when the traffic is beyond the breakdown point (i.e. point of critical flow), if traffic starts to recover (i.e. vehicle density along the study segment starts decreasing), flow does not reach its original maximum value (for each density value), that phenomenon is called hysteresis. Figures 1-a and 1-b illustrate the concepts of critical density and hysteresis. These figures were created by simulating human-driven vehicles within a loop environment which we utilize for the purposes of this study. As seen in Figure 1-b, after the breakdown point, by reducing the density, the flow doesn't reach its previous levels. The main contribution of this study is to propose a framework that utilizes deep reinforcement learning in the presence of CAVs under various market penetration rates (MPR) to specifically tackle the issue of traffic recovery beyond the point of critical density. This framework aims to prevent hysteresis formation and can often increase the flow level to the breakdown point level, significantly improving the efficiency of the system.

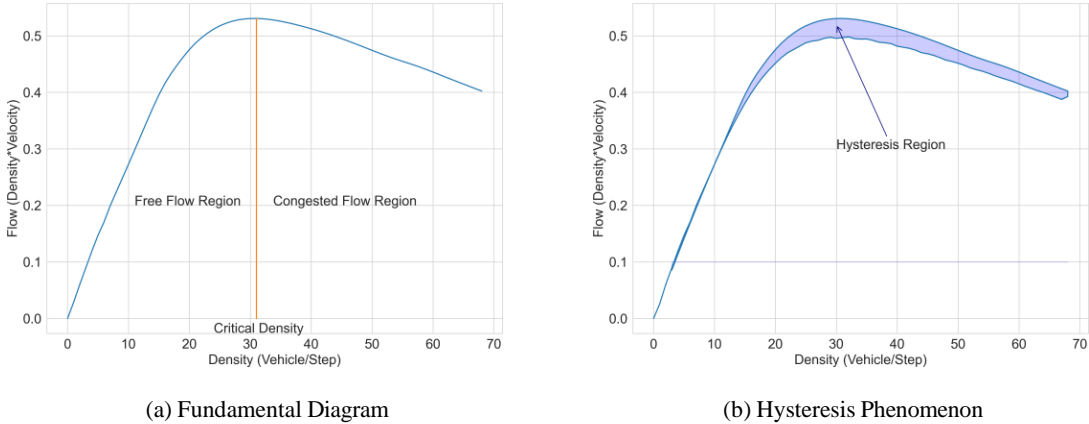

(a) Fundamental Diagram

(b) Hysteresis Phenomenon

Figure 1: (a) Fundamental Diagram During the Loading Phase, and (b) Hysteresis Phenomenon During the Unloading Phase. [38]

The rest of this paper is structured in the following order: the next section discusses the formulation and details of the proposed Deep Reinforcement Learning (DRL) model. This is followed by an introduction to the simulation setup and model parameters. Subsequently, detailed results and a discussion on the findings of this paper are presented. And finally, we conclude the paper with a brief summary and a look at possible future research directions.

## 2   Model Formulation

### 2.1   Double Deep Q Network

The network architecture chosen in this work is in the form of Double Deep Q Network (DDQN) [39]. This framework was selected because it provides a simplified structure through which the concept we are trying to evaluate can be directly tested and verified. Note that this framework has been successfully utilized in many studies, including our own previous work [40]. The Double Deep Q Network is a variation of the Deep Q Network where a reinforcement learning method called Q-learning is integrated with deep neural networks to perform end-to-end control of the agent(s). Two neural networks are present in both the DQN and DDQN architectures: main and target networks. The main network takes the defined states of the environment as inputs and outputs one of the possible actions at any given state. As in Q-learning, the famous Bellman equation [41] is used to update the weights of the network and the Q values. The target network is a network similar to the main network but is occasionally updated rather than at every step in order to increase the stability of the model and attain better convergence rates. In the DDQN, overestimation is reduced by decomposing the max operation in the target network into action selection and action evaluation. This addition further enhances the performance of the model [42, 41, 39]. Figure 2 illustrates a simple depiction of a Deep Q Network where actions are taken by the agent, which results in new states and rewards observed from the environment, and finally, those new states and rewards are fed into the neural network (i.e. agent) again to choose a new action.

### 2.2   Multi-Agent Structure

In multi-agent reinforcement learning, learning and execution can be categorized into 3 main categories; centralized training-centralized execution, centralized training-decentralized execution, and finally decentralized training-decentralized execution. In the first category, the agents in the environment are treated as a single entity and can share one learning model (i.e. centralized controller). The centralized controller determines the actions of each agent, which is updated using the inputs and rewards from all agents simultaneously. Note that the model can output actions to be taken by all agents uniformly or it can output a different action for different agents. On the other hand, centralized training-decentralized execution means that the model takes inputs from all agents to tune its parameters, however, unique actions are taken by each agent independently based on the local states that the agent observes. Finally, decentralized training-decentralized execution entails the structure where each agent has a separate model that is trained using the inputs of that specific agent, and similarly, each agent executes actions as instructed by its respective model.

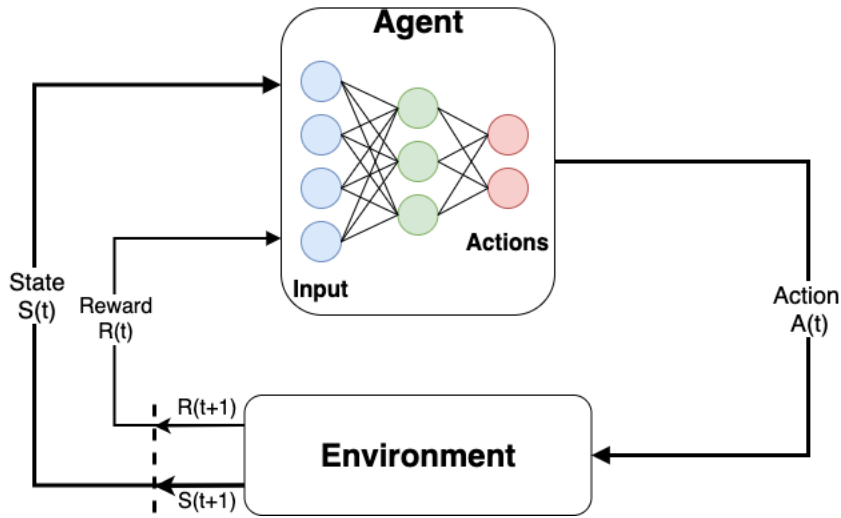

Figure 2: Simple Illustration of DQN [41]

The agents, however, usually keep some connection to the outcome at the centralized level to ensure a desirable outcome. Naturally, there are advantages and disadvantages to each of the aforementioned methods. For instance, in centralized training-centralized execution, the centralized decision-making process can result in coordinated and effective learning [43]. Note that if only one model is utilized (uniform actions among all the agents), the complexity and computational requirements of training and execution can be reduced significantly. However, the assumption that all agents should take the same action regardless of their different states and observations is rather simplistic. Due to the complexity of the decision space, centralized training theoretically leads to higher variance in policy updates [44]. On the other hand, lower variance in policy updates can be achieved by increasing the complexity by defining several models and issuing separate actions for the agents depending on their own observations, i.e., decentralized training-decentralized execution. In practice, decentralized training regularly gives more robust performance [44]. However, a shortcoming is that in a decentralized algorithm, the credit assignment system among agents (i.e., who gets the credit for a job well done?) is challenging to tune and can often lead to miscoordination among agents [45]. In this work, we adopt the somewhat more simplistic structure of centralized training-centralized execution where a model is jointly trained by the states of all agents (i.e. CAVs) on the road, and a central controller issues a single action at every time step that needs to be executed by all CAVs uniformly. As discussed earlier, there are shortcomings to this structure. Most importantly, it is obvious that each CAV would be better off executing its own action (i.e. acceleration/deceleration) based on its local conditions (i.e. proximity to leader and follower), rather than follow a global command that is to be executed by all. However, we demonstrate the ability of this model to successfully shape the traffic state and influence flow positively. It is important to note that such an approach becomes less effective when CAVs are randomly distributed in the traffic stream.

### 2.3 Model Parameters

To define our agents, we create a fully connected deep neural network with 4 hidden layers. The first two layers consist of 512 neurons while the 3rd and 4th layers have 128 and 64 neurons, respectively. Figure 3 shows the details of the network architecture. The fully connected neural network was able to capture the complexities of the task and performed very well with a reasonable amount of training time. It is noted that multiple network structures were experimented with along with experimentation on other model parameters before choosing the final network structure. Table 1 presents the hyperparameter setting for the formulated DDQN model. Throughout the training and testing procedures, we experimented with discount factor ($\gamma$) values ranging between 0.75 and 0.999. The final value selected, i.e. 0.90, ensured the RL agent gives more weight to current rewards rather than to future rewards when making a decision ($\gamma = 1.0$ means that the agent considers no difference between the current reward and future reward, i.e. the agent becomes more farsighted [41]). This is done since no task-finishing 'success' reward was required for this problem formulation. In another case where a success reward may be needed, a higher discount factor may be used to give a large enough consideration to this final reward that may occur after many time steps in the episode. Moreover, we choose a replay memory [46] size of 100,000 to limit the computational load during training. We also choose a window length of 1. The window size controls how many steps in time are taken into consideration as input into the neural network (i.e. features). Choosing a window length larger than 1 is advantageous in scenarios where actions taken at any time step are dependent on the previous state of the system. On the other hand, choosing a window size of 1 means that actions can be chosen accurately enough by only observing the current states. We train the agent for 1 million steps to achieve a good level of convergence. During the training process, we select the epsilon-greedy policy [41] with a linearly decaying epsilon starting at 1 and ending at 0.05. This ensures the agent explores for an adequate amount of time before starting to follow greedy action choices. This guarantees sufficient exploration of the environment is attained and thus a favorable/near-optimal performance is achieved. Note that minimum $\epsilon$ is set to 0.05 to ensure some level of exploration throughout the calibration process. This is critical to ensure that the system does not stay within a local minimum. Finally, we add a time-step learning rate decay schedule where the learning rate decreases as we progress further in the training. This helps the model converge in a more stable fashion when compared to a fixed learning rate.

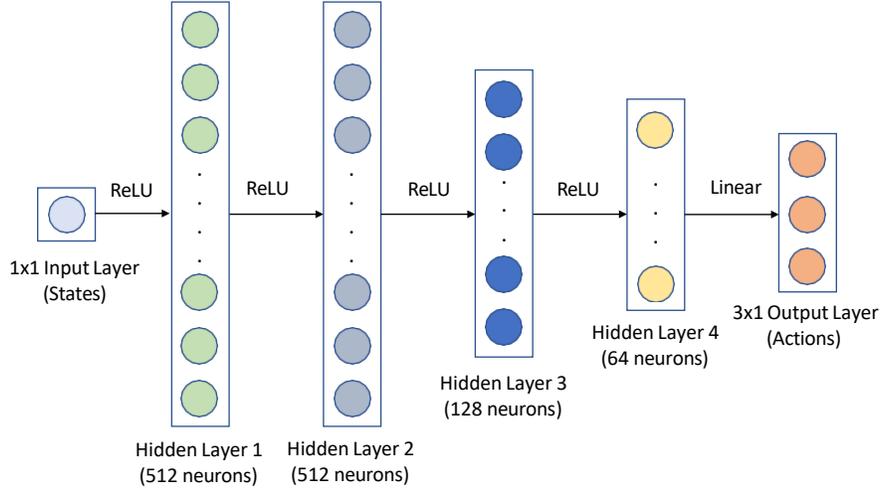

Figure 3: Study's Deep Q Network Architecture

Table 1: DQN Hyperparameter Setting

| Hyperparameter | Value |
|---|---|
| Number of Layers | 4 |
| Number of Hidden Units | 512, 512, 128, 64 |
| Learning Rate | 0.001 → 0 |
| Policy | Epsilon-Greedy |
| $\epsilon$ | 1.0 → 0.05 |
| Discount Factor $\gamma$ | 0.90 |
| Replay Memory Size | 100,000 |
| Number of Episodes | 5,000 |
| Batch Size | 32 |
| Activation Functions | ReLU, Linear |
| Optimizer | Adam |

## 3 Simulation Set-up

### 3.1 Simulation Environment

In order to simulate the events of traffic flow and create the fundamental diagram of traffic flow, we create a single-lane loop system. A loop system was adopted for its ability to simulate the events of congestion and hysteresis. While the dynamics of a ring road can be different from a straight road, it provides all the necessary features to evaluate the main hypothesis of this study and it is significantly easier to create congestion in such a structure. Note that a single shockwave can potentially propagate infinitely in a loop system. A simplified depiction of the loop system is shown in figure 4. The loop begins with no vehicles inside it and they begin to show up and density increases incrementally with time. Human-driven vehicles are modeled using the Intelligent Driver Model (IDM) [47]. All vehicles in the first phase of the simulation are human-driven. The system continues until reaching the critical density and then continues beyond traffic breakdown. The total length of the loop is designed to be 1000 m, and the number of vehicles needed to reach the desired congestion level is approximately 68 vehicles (note that this number can change by changing the loop length). After reaching this density point, a shock to the system is induced by removing a portion of the vehicles in order to observe the response. Figure 5 shows the response of the system if vehicles are removed incrementally. In this case, the vehicles are removed in a random manner. The results of several simulations with random removal showed that the response of the system remained the same. It can be seen that the flow never reaches its original maximum value until almost no vehicles are left in the loop (i.e., hysteresis).

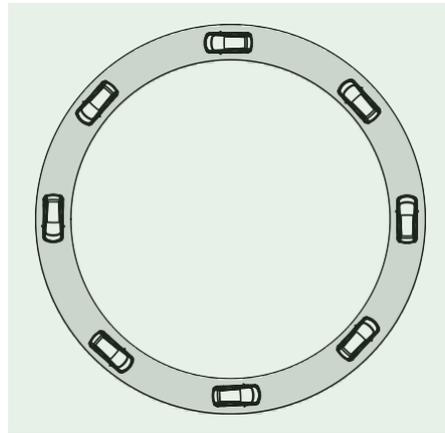

Figure 4: Loop Environment

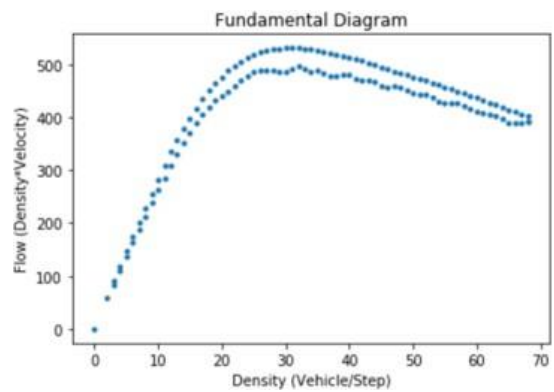

Figure 5: Hysteresis Formed By Human-Driven Vehicles

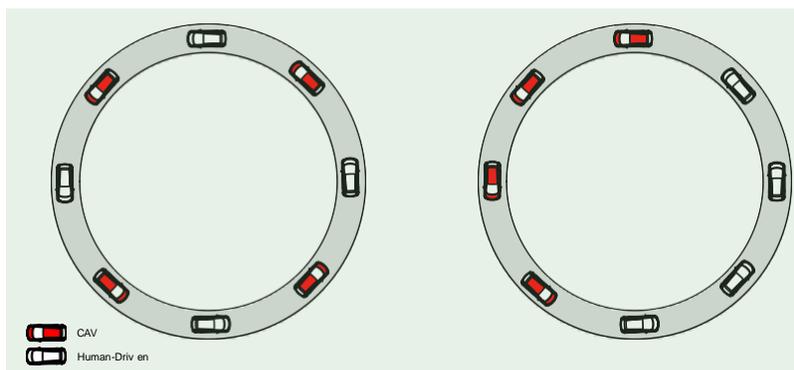

Figure 6: CAVs General Formation Strategies

Alternatively, a portion of the human-driven vehicles is substituted with CAVs which are governed by centralized action selection through the RL model defined earlier and are assumed to have full communications with the infrastructure. When the shock is introduced to the system, the task of the RL agents is to recover traffic flow to values higher than that of the original maximum observed value (for a specific density). It is important to note that the instant substitution of human-driven vehicles with CAVs may introduce some issues in practice, however, the proposed system is expected to handle a slight delay in switching (similar to the delay when vehicles switch to adaptive cruise control). Regardless, the focus of this study is not on evaluating the feasibility of such a change nor on driver compliance (although both are critical and determining factors).

The RL agents are assumed to be uniformly distributed among the vehicles within the loop. The uniform formation of the CAVs among the vehicles was found to yield favorable results and is adopted. Note that while this can be considered a limiting assumption, for any required penetration of RL agents, a close-enough formation (to uniformly distributed formation) can be achieved by utilizing only a subset of CAVs, assuming that a minimum number of CAVs is available and they are randomly distributed in the traffic. Obviously, the required minimum number of CAVs can change based on the desired penetration of RL agents. There is, however, one particular case that cannot be achieved with this method (except at very high market penetration rates of CAVs): RL agents forming a platoon. Accordingly, both of these formations were tested. Figure 6 illustrates the aformentioned formation strategies. Note that the effects of RL vehicles on traffic flow reduces when multiple RL vehicles follow each other. Therefore, the dynamics of the system with slight deviation from uniformly distributed RL agents can be different from the cases persented in this study. Since in a real-world setting, we will have little control on the availability of the RL vehicles (due to several factors, including their location and driver compliance), it is essential to design a control system capable of handling any formation of RL vehicles. Accordingly, this study is considered a proof-of-concept for the proposed control strategy and the generalized case has been left for future research. We experiment with different market penetration rates of RL agents (ranging between 15 and 66 percent) to test the capability of the CAVs in influencing the system. We also provide a comparison between this framework and other popular existing congestion management strategies.

### 3.2 Reinforcement Learning Environment

The task of the CAVs is to recover from the loss of flow after breakdown. Recovery is measured after a portion of vehicles leave the loop and the RL vehicles are tasked with increasing the flow level to a desirable value (if possible), i.e., preventing hysteresis in the fundamental diagram. The CAVs are assumed to maintain full communication with the infrastructure and have a few sensors such as Inertial Measurement Units (IMU), GPS, and perception sensors (e.g., LiDAR or radar) through which they can measure the approximate speed of the vehicles ahead and behind (which will be used by the system to estimate the speed of the system). The average speed of the system is then utilized as the state (input) to the deep reinforcement learning (DRL) model. On the other hand, the action space from which the CAVs choose their actions is defined as acceleration/deceleration values ranging between -1, 0, and 1 $m/s^2$. Note that this is a shortcoming of the DDQN algorithm where the action space is discrete and as such, we limit our action space to three values, this is also why the number of neurons of the output layer in the DQN architecture is set to three (corresponding to the three possible actions). Moreover, more aggressive acceleration/deceleration values can be chosen which could result in reaching the desired flow much faster, however, at the cost of passenger comfort. This definition of state-action space allows the DRL to speed up/slow down the CAVs depending on the given average speed and ultimately alter the flow of the whole system. Subsequently, we define a reward system that maximizes the reward of increased flow and penalizes the event of a collision in the system. Increasing the flow is rewarded based on the following equation:

$$r(v) = v_a \tag{1}$$

where,

- r(v) is the reward
- $v_a$ is the average speed of the system

The reward in equation 1 is calculated every time step during an episode. Accordingly, the total episode reward is maximized if the average speed of the system is maximized throughout the episode. Alternatively, a below-optimal reward would be achieved if the average speed of the vehicles remains below the maximum attainable speed. The other element of reward is the negative reward associated with a collision. This is a once-per-episode type of penalty that occurs if any vehicle in the loop collides with any other. In the event of a collision, the whole episode would terminate. A final success reward is defined to drive the RL model to converge to the desired solution. 'Success' here is defined when the flow passes the original maximum flow value defined by the fundamental diagram during the loading process. This helps the model not get stuck at local minima where it is accumulating rewards but not to the point that is actually desired by the system. As will be shown in the next sections, such an architecture can be used for traffic shaping and achieving a desired flow and speed. In other words, rather than setting the desirable flow to be beating the original maximum flow, we can set it to be a certain flow value, and train the model to regulate the system to that level of flow. This has several usages ranging from traffic shaping, speed harmonization, collision mitigation, and others. Table 2 presents a detailed overview of the reinforcement learning and road environment parameters.

Table 2: Reinforcement Learning and Road Environment Setting

| Parameter | Value |
| --- | --- |
| Number of States | 1 |
| Number of Actions | 3 |
| Collision Reward | -3000 |
| Average Speed Reward | System Average Speed/step |
| Success Reward | +1000 |
| Acceleration/Deceleration Values (m/$s^2$) | -1, 0, 1 |
| Desired Velocity (m/s) | 30 |
| Time Step Length (s) | 0.1 |

It is noteworthy that different reward elements were tested and the presented structure was selected based on a trial and error process. This specific combination of reward values was among the best performing for our specific problem. However, this structure is not unique and other reward structures did perform the task successfully. On the other hand, it is important to keep in mind that a certain change in the scenario or the environment may require a change in the reward structure or the learning procedure, therefore, in practice, it is essential to build robust models that are able to adapt to different scenarios and scale beyond a simulation environment. This is generally a weakness of RL-based approaches as they can be very sensitive to the reward structure.

## 4 Results and Discussion

The first step before testing the proposed structure is to infer the minimum MRP required to be able to influence the system. For instance, one CAV would not be able to influence the state of a fleet of over 60 vehicles. Below we discuss this approach.

Assume an existing density of vehicles on a certain road segment ($k_1$), with an existing average time headway $\bar{h}_{k_1}$, and existing flow $q_{k_1}$. If a decrease in density occurs while maintaining the same speed for the system (a sudden drop in density), flow would decrease directly (give $q = ku$. Given, that flow ($q$) is inversely proportional to headway ($h$), in such a case, decreasing the time headway after the departure of some vehicles (i.e. maintaining the headway pre-departure) would result in maintaining the flow of the system. As such, the following equation can be used to calculate the required number of CAVs to maintain the flow of the system.

$$\bar{h}_{t-1} = \frac{(NT_t - NN_t) \cdot \bar{h}_t + h_{CAV} \cdot NN_t}{NT_t} \quad (2)$$

where,

- $\bar{h}_{t-1}$ is the average time headway at the previous time step
- $\bar{h}_t$ is the average time headway at the current time step
- $NT_t$ is the total number of vehicles at time t
- $NN_t$ is the number of CAVs needed to maintain $\bar{h}_{t-1}$
- $h_{CAV}$ is the desired time headway of the CAVs

Given equation 2 above, solving for the total number of CAVs required to maintain the previous time headway ($NN_T$) gives us the minimum MPR required to have an effective framework. As such, given an example case where there are 68 vehicles in total, a current time headway of 2.549 seconds, and a desired CAV time headway of 2 seconds, if 1 vehicle departs the system, i.e., $NT_t$ becomes 67 and current time headway becomes 2.5779, we conclude that 5 CAVs would be needed to maintain the previous time headway, and in turn, maintain the flow. As mentioned earlier, the average time headway ($h$) can be inferred from the average flow of the system ($\bar{q}$) using the inverse relationship ($h = \frac{1}{\bar{q}}$).

We test the ability of our framework with three main MPRs; 15%, 33%, and 66%. Additionally, we test the ability of the model to perform the task of traffic flow shaping through 2 episodes of traffic density shifts. First, we show the performance of the IDM model as a baseline. In this trial, all vehicles on the loop are human-driven (modeled using IDM). Figure 7 below shows the performance of the human-simulating IDM model in recovering traffic flow after some vehicles depart the loop. It can be seen in Figure 7-a that that flow never recovers to the state at which it was at during phase 1 (i.e. on the fundamental diagram). It can also be seen that the maximum speed attained plateaus at around 9 m/s.

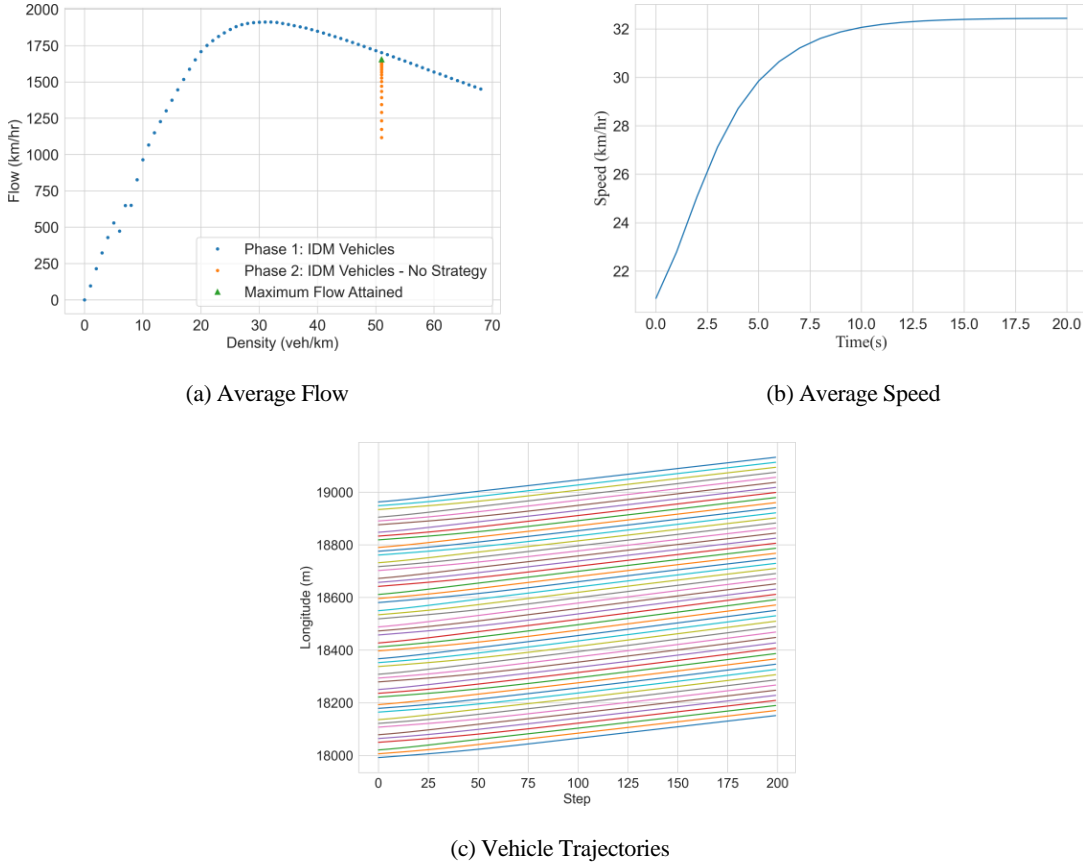

Figure 7: The Performance of the System with 100% IDM Vehicles.

Conversely, we perform the same experiment but at the beginning of the second phase (i.e. after some vehicles depart the loop), we switch 33.3% of the vehicles to be CAVs. The total number of vehicles on the loop before any departure happens is 68 vehicles. 17 vehicles depart the loop leaving 51 vehicles inside the loop. Of those 51 vehicles, 17 vehicles are CAVs and are spread uniformly along the loop. Figure 8 illustrates the performance of the system with 33.3% CAVs present within the loop. It can be seen that the DRL-trained CAVs are able to substantially influence the state of the system. Figure 8-a shows how the flow does not just match the original flow represented by the point on the fundamental diagram, but far exceeds it. The maximum average speed attained increases to 11 m/s, which is more than a 22% increase when compared to the 9 m/s of the 100% human-driven system. Vehicle trajectories in Figure 8-c also show the ability of the CAVs to utilize the gap created by the departing vehicles and create and maintain the improved traffic state. Figure 8-d shows the trend of reward during the training process of the DDQN model. While some dips in the reward occur during training - which are caused by collisions happening as the agents progress in their learning process - a steady increase in the reward is clear, which nearly plateaus towards the end.

It is worth noting here that for the results of this work, the maximum reward values and the number of episodes required, vary across different scenarios. This is due to the fact that convergence (i.e. stopping criteria) was set differently for each task because of the changes in each scenario. For instance, the task of improving flow after the first step of density change differs from the task of the second scenario where the density is decreased a second time from a different starting point. Accordingly, to ensure a valid comparison, success is measured in terms of achieving the actual goal rather than reaching a certain reward value or a number of steps.

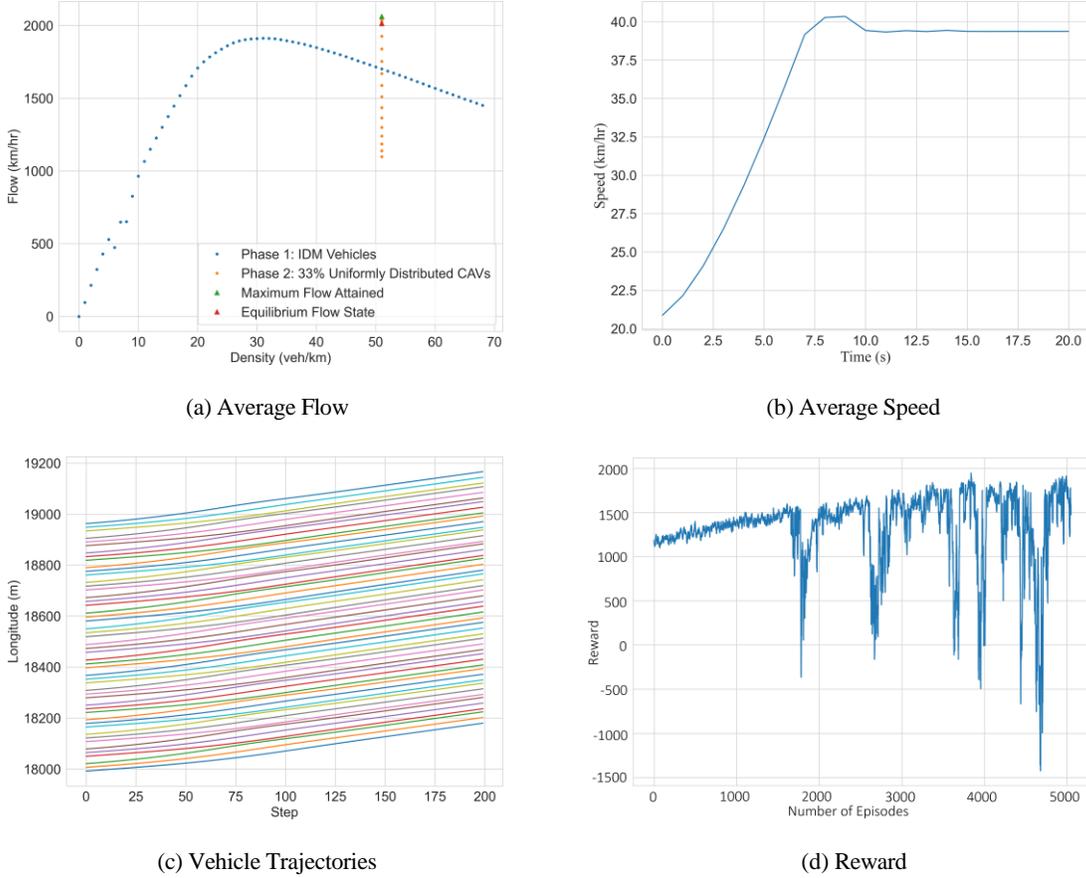

Figure 8: The Performance of the System with 33.3% RL-Controlled CAVs

Subsequently, we test the ability of the proposed framework to influence traffic flow at smaller and larger MPRs, i.e., 15% and 66%. Accordingly, for the 15% case, 9 vehicles out of the 68 vehicles, are removed from the loop, leaving 59 vehicles in the loop with 9 of those being CAVs. This amounts to 15.25% MPR, to be precise. Figure 9 shows the performance of the framework with 15% CAVs in the loop. Figure 9-a shows the ability of the 9 CAV vehicles to impact the system to a great extent. The system is able to surpass its original flow by a good margin. Given that in this scenario the number of remaining vehicles on the loop is 59, the maximum attainable speed is lower than that of the previous scenario (i.e., 33%). Figure 9-b shows that the maximum average speed that was achieved is around 8 m/s. Vehicle trajectories and reward figures in Figures 9-c and d are consistent with that of the previous scenario, where CAVs are utilizing the gap to increase the flow of the system and the learning process is converging towards the end of the training episodes.

Alternatively, figure 10 shows the performance of the system under the 66% MPR. Similar to the first case of 33% MPR, 17 vehicles depart the loop, however, 34 vehicles are subsequently turned into CAVs. It is shown that with the increase in market penetration rate, the system is able to achieve the goal more efficiently.

Additionally, we further explore the potential of this framework by adding a second step to the 33.3% MPR scenario. This second step is defined as another phase of vehicle departure from the loop which is then followed by the attempt of the CAVs to improve the state of the traffic again. For this scenario, after the departure of the first 17 vehicles, 12 more vehicles depart. This leaves 39 vehicles on the loop, 13 of which (33.3%) are CAVs. We then re-train the DDQN model to optimize the output. This scenario is a proof of concept that the proposed framework is able to improve the overall traffic state given any initial settings, and can continue to do so as traffic density changes. It can be seen in Figures 11-a through d that the model is able to converge to an optimal solution, where the flow is maximized and in fact, also surpasses that of the original state. Average speeds reach values above 12 m/s, and interestingly enough, the trajectories show that, while speeds and flow are still increasing, the headway between the vehicles within the loop is reaching equilibrium.

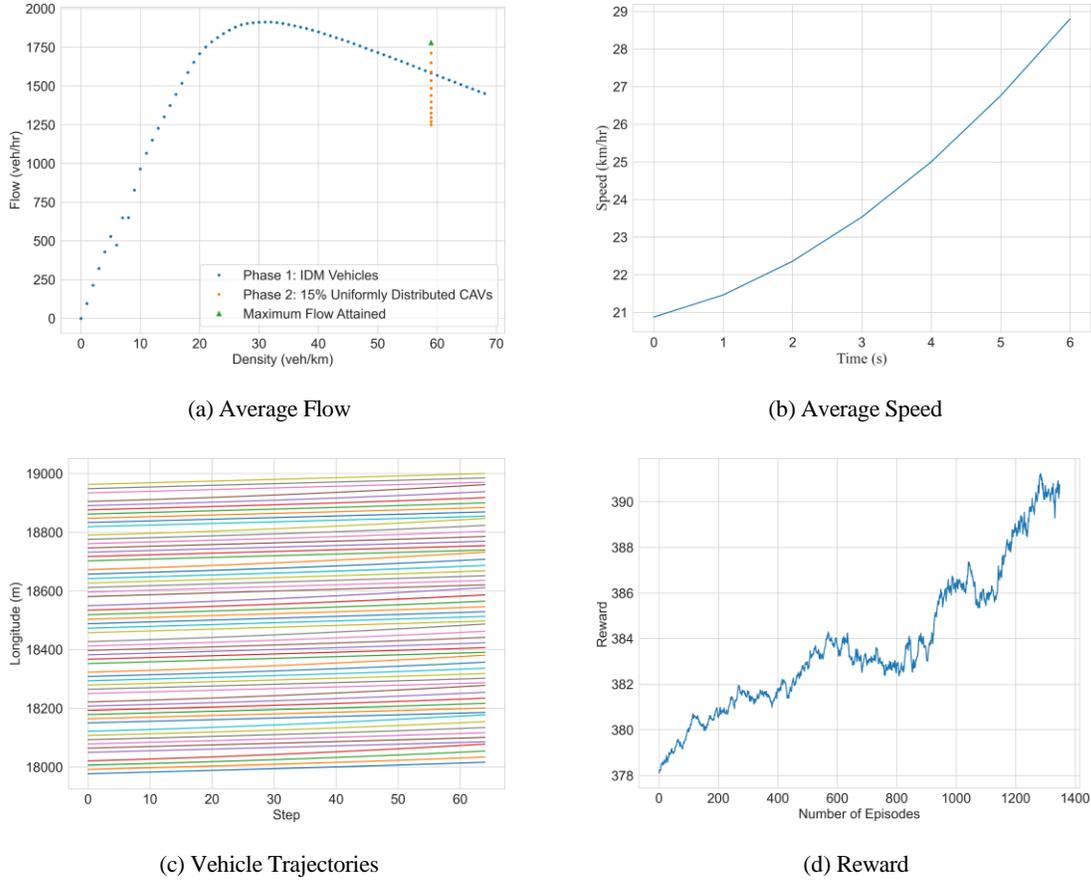

Figure 9: The Performance of the System with 15% RL-Controlled CAVs

On the other hand, we explore the ability of the CAVs to maintain the flow that they achieve and that the positive effect is not instantaneous and then vanishes. To do so, we introduce a twist to the first scenario. Instead of keeping the CAVs as they are after reaching the maximum flow attained, we turn them again to human-driven vehicles and run the simulation for an extra 200 steps from that point. Given that the time step is 0.1 seconds, the total extra time added is 20 seconds. Figure 12 shows a comparison between running each mode for 20 extra seconds as discussed. It is clear to see that the CAVs not only reach this desirable high level of flow illustrated before, but they are also able to maintain a very similar level of flow for a significant amount of time. On the other hand, it can be seen in Figure 12-b that when switching the vehicles back to the IDM model (i.e. to be human-driven), the flow that was attained by the CAVs quickly diminishes and the traffic state goes back to a level substantially lower than it was 20 seconds ago.

Finally, to illustrate the advantage of this framework against other congestion management strategies, we test a popular speed harmonization strategy which was proposed by Allaby et al. [48] and has been utilized by many other studies ([49, 50, 51]). The strategy is enforced on all vehicles in the loop and is contrasted with a 100% human-driven (i.e. IDM) system. Figure 13 shows the results of the experiment. It can be seen that the speed harmonization strategy (as may be expected) worsens the ability of the system to recover after traffic breakdown rather than improving it. This is expected as speed harmonization tends to reduce the speed limit, which reduces the maximum achievable flow. Such a congestion management system is only effective before breakdown formation to prevent flow from reaching the critical level. A similar argument can be made for ramp-metering, except ramp metering limits the density instead of speed. Note that ramp-metering is not tested in this paper due to the structure of the loop system.

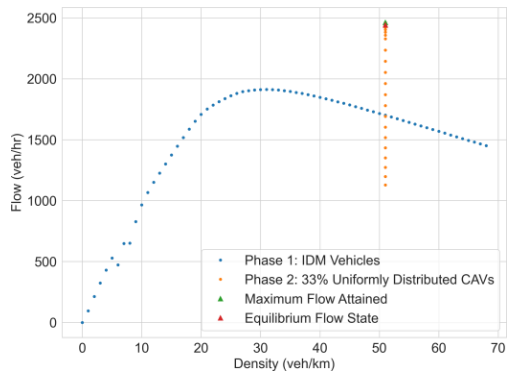
(a) Average Flow

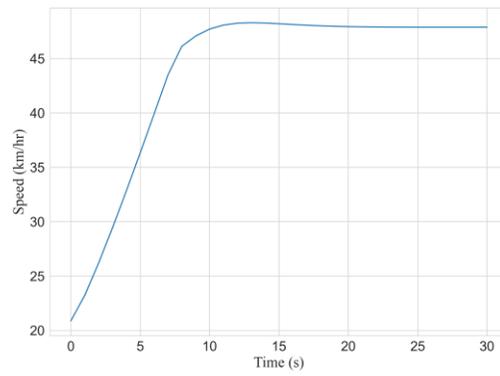
(b) Average Speed

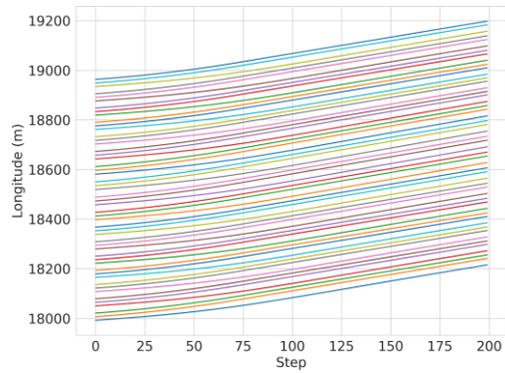
(c) Vehicle Trajectories

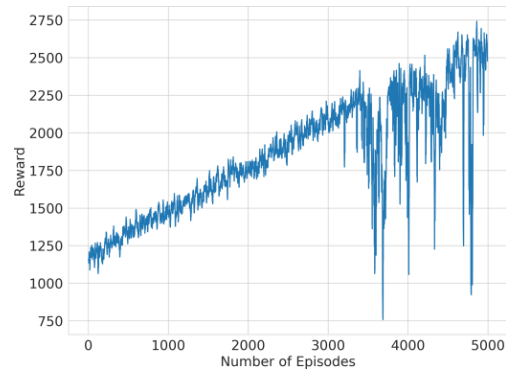
(d) Reward

Figure 10: The Performance of the System with 66% RL-Controlled CAVs

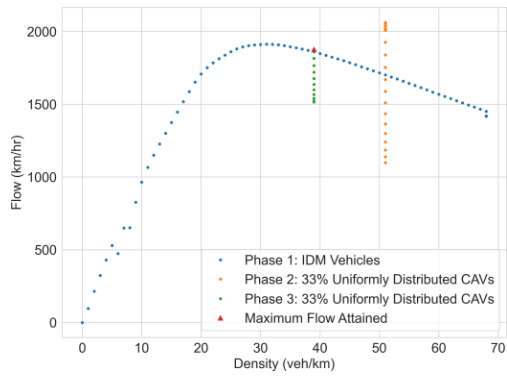
(a) Average Flow

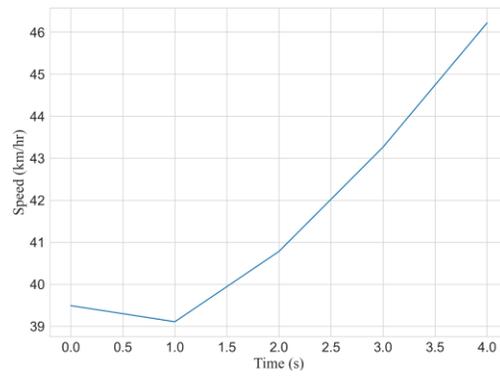
(b) Average Speed

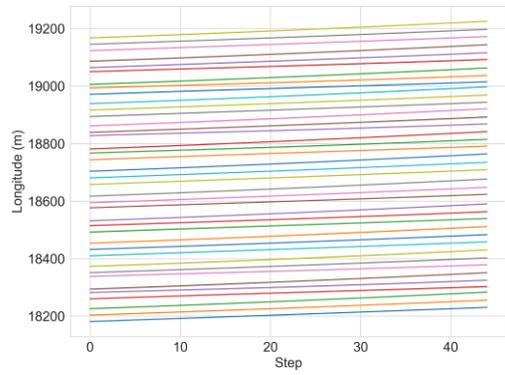
(c) Vehicle Trajectories

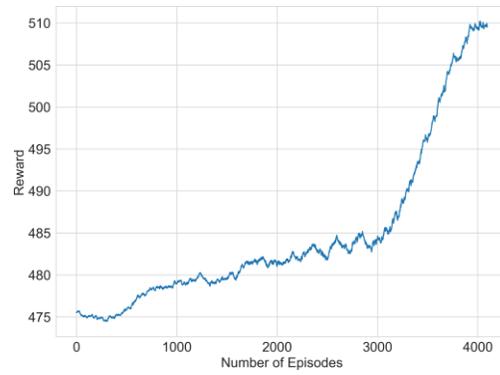
(d) Reward

Figure 11: 33.3% CAVs Performance - Second Scenario

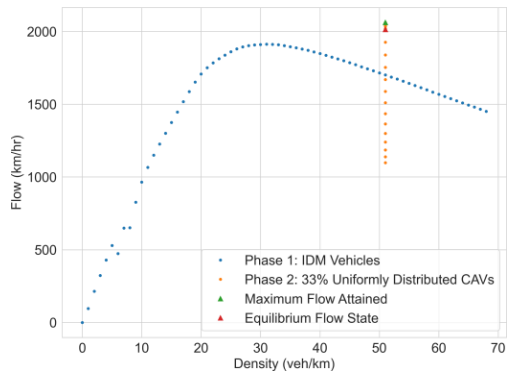
(a) Average Flow - CAVs

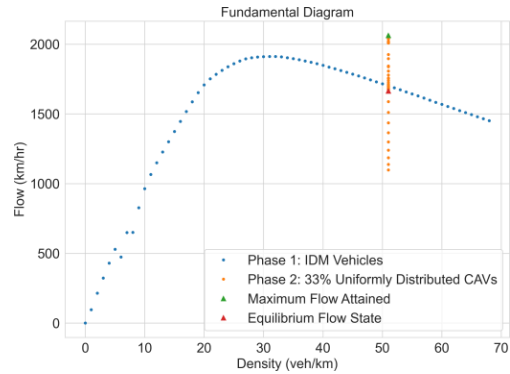
(b) Average Flow - IDM

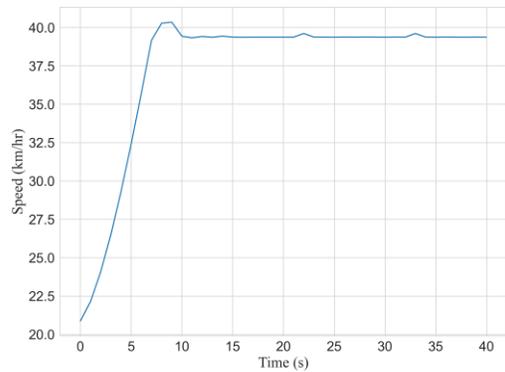
(c) Average Speed - CAVs

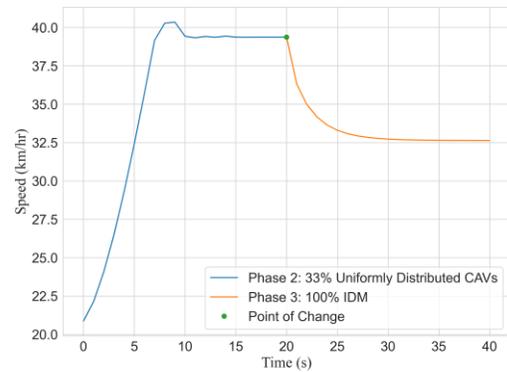
(d) Average Speed - IDM

Figure 12: CAV vs IDM Comparison

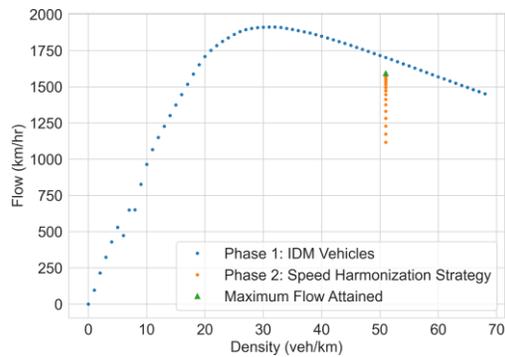
(a) Average Flow - Speed Harmonization Strategy

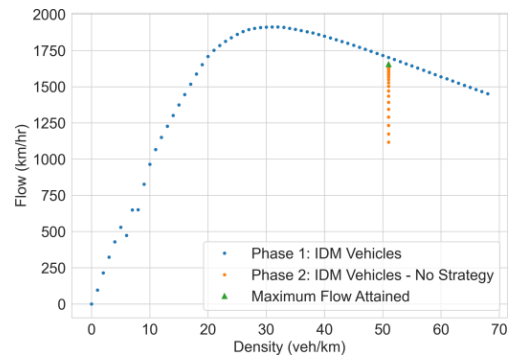
(b) Average Flow - IDM

Figure 13: Speed Harmonization vs IDM Comparison

# 5 Conclusion and Future Work

This study proposes a framework with the primary goal of traffic shaping and congestion mitigation. To this end, the framework utilizes deep reinforcement learning and a formation strategy to optimize the behavior of Connected and Automated Vehicles (CAVs) present within a loop road environment. The purpose of the CAVs is to control traffic flow in a way that improves the overall traffic condition. The Double Deep Q Network (DDQN) is chosen to train our agents (i.e. CAVs) to perform the task at hand, where the input state to this network is the average speed of the system at every time step, and the corresponding possible output actions are acceleration/deceleration values consisting of -1, 0, and 1 $m/s^2$ for the agents to choose from. The proposed method shows significant improvement with this discretized action space and is expected to perform even better with a continuous action space (which, at this point, has been left for future research). This problem is treated as a centralized training, centralized execution task, where centralized training is defined by the one model that is trained using the average speed of the traffic, and centralized execution refers to the singular action that is transmitted by the model to be taken by all CAVs on the loop. This is a simplified version of the more complex decentralized training, decentralized execution approach that can be utilized in this environment. Using a ring road simulation, we show the ability of the framework to significantly improve traffic flow after it had reached a state beyond the point of critical density, and that the CAVs are capable of shaping traffic and achieving a desired flow level (given a minimum MPR of CAVs). We also compare this framework to popular congestion management strategies and illustrate its superiority. The framework assumes the CAVs possess some level of sensory ability allowing them to infer the average speed of traffic which is a relatively simple task given that CAVs have some common sensors including GPS and radar. For future work, the cases of decentralized training-decentralized execution, as well as centralized training with decentralized execution will be explored for the purpose of traffic shaping. Exploring different deep reinforcement learning models and their performance will be required and updated definitions of states and actions will be investigated.

# 6 Acknowledgments

This material is based upon work supported by the National Science Foundation under Grant No. 2047937.

# 7 AUTHOR CONTRIBUTION STATEMENT

The authors confirm equal contribution in studying conception and design, analysis and interpretation of results, and manuscript preparation. All authors reviewed the results and approved the final version of the manuscript.